\documentstyle[prc,aps,multicol]{revtex}

\begin{document}

\draft

\title{ Spectroscopy of $^{13,14}$B via the one-neutron knockout reaction.}

\author{ V. Guimar\~aes and J. J. Kolata}
\address{ Physics Department, University of Notre Dame, Notre Dame,
IN 46556 U.S.A}

\author{D. Bazin, B. Blank, B. A. Brown, T. Glasmacher, P. G. Hansen,
R. W. Ibbotson, D. Karnes,V. Maddalena, A. Navin, B. Pritychenko, and B. M. Sherrill}
\address{ National Superconducting Cyclotron Laboratory,
Michigan State University, East Lansing, MI 48824, U.S.A}

\author{D.P. Balamuth and J.E. Bush}
\address{ Physics Department, University of Pennsylvania, Philadelphia,
PA 19104 U.S.A}

\date{\today}

\maketitle

\begin{abstract}
The single-nucleon knockout reactions $^9$Be($^{14}$B,$^{13}$B+$\gamma$)X
and $^{197}$Au($^{14}$B,$^{13}$B+$\gamma$)X, at an incident energy
of 60 MeV per nucleon, have been used to probe the structure of
$^{14}$B and of the core fragment $^{13}$B.  A dominant {\it 2s}
configuration is deduced for the neutron in the ground state of
$^{14}$B.  The longitudinal momentum distribution for this state
is consistent with ``neutron halo" structure.  Spin assignments for
$^{13}$B excited states at 3.48 and 3.68 MeV are proposed based
on the observed spectroscopic factors for one-neutron removal.
\end{abstract}


\begin{multicols}{2}

\section{Introduction}
The structure and reactions of ``neutron-halo" nuclei, weakly bound systems
that display a very diffuse surface of nearly pure neutron matter at
densities far below that of normal nuclear matter, have recently been
subjects of intense study \cite{han95}.
In a highly simplified picture, the longitudinal momentum distribution of
fragments from
the breakup of a loosely bound projectile directly reflects the internal
momentum distribution of the valence nucleon and hence the square of 
the Fourier transform of its wave
function. Thus, halo formation in such loosely bound nuclei can
be investigated by measuring the momentum distribution
of the fragment from a breakup reaction.
The wide spatial dispersion characteristic of a halo neutron translates
into a narrow momentum distribution.  In this model, one treats the
core of the nucleus as an inert spectator to the breakup process.
While early experiments were based on the assumption that the measured
momentum distribution represented a single reaction channel, it has recently become
possible to go beyond this naive
approach by measuring $\gamma$-radiation from the decay of excited
states of the core in coincidence with the breakup fragments \cite{nav98}.
This technique provides a further benefit in that partial cross
sections for the excitation of different final states of the core
can be measured, together with the longitudinal momentum distributions
associated with inert and `active' cores.  This allows for spectroscopic
investigation of both the core nucleus and the valence nucleon(s).
The method, originally developed in a search for proton halo nuclei
\cite{nav98}, has also been applied to the neutron case \cite{aum99,mad00}.

Among the candidates for neutron halo formation, the $^{14}$B nucleus is of
particular interest as the lowest-mass bound system amongst the N=9 isotones.
The well-known halo nuclei $^{11}$Be, $^{11}$Li, $^{14}$Be, $^{17}$B, and
$^{19}$C occupy a similar position for N=7, N=8, N=10, N=12, and
N=13, respectively (see Refs. \cite{sme99,nak99,bau98} and references therein 
for a discussion of $^{19}$C structure).
Moreover, $^{14}$B is an odd-odd nucleus and thus different from any other 
neutron-halo system observed to date.  In the first attempt to probe the structure of 
$^{14}$B, Bazin, {\it et al.}\cite{baz98} reported a broadening of the longitudinal 
momentum distribution when compared with the shape expected from theoretical
predictions, and
suggested that this difference might be due to a strong contribution from core
excitation. As a result, they were unable to draw any firm conclusion about
halo
formation in this nucleus.  In order to investigate the structure of
$^{14}$B, and
to determine the role that the core plays in the dissociation reaction,
the momentum distributions of $^{13}$B fragments corresponding to removal
of neutrons from different orbitals in $^{14}$B were measured using the
technique
described in Ref.\cite{nav98}.
The contribution from core excitation was isolated
by measuring the momentum distribution of excited $^{13}$B
fragments in coincidence with $\gamma$-rays from their decay.

This paper is divided into the following sections:
the experimental setup and the procedure are described in
Section II, while the experimental results and
the analysis of the longitudinal momentum distributions is presented in
Section III. Finally, a summary is given in Section IV.

\section{Experiment}

           The experiment was performed at the National Superconducting
Cyclotron Laboratory (NSCL) at Michigan State University.
The $^{14}$B  radioactive secondary beam was produced  by  fragmentation
of an 80 MeV per nucleon primary $^{18}$O beam on a 790 mg/cm$^2$ Be target.
The secondary $^{14}$B beam, having an energy of 59.2 $\pm$ 0.3  MeV per
nucleon, was then selected by the A1200 fragment separator \cite{she91}
and transmitted to a target chamber where a neutron was removed during
an interaction with a Be or Au target, leaving
the $^{13}$B core in its ground state or in an excited state.
The $\gamma$-rays from the decay of $^{13}$B excited states
were detected by an array of 38 cylindrical NaI (Tl) detectors \cite{sch99}
which completely surrounded the target.  Recoiling $^{13}$B core fragments
were momentum-analyzed using the S800
spectrograph \cite{she99}. The momentum acceptance of the spectrograph
was 6$\%$ and the angular acceptances were $\pm 3.5^{\rm o}$ and
$\pm 5^{\rm o}$ in the dispersive and non-dispersive directions,
respectively. These acceptances resulted in an estimated 99$\%$
efficiency for detecting events corresponding to the ground-state momentum
distribution, and 95$\%$ for the momentum distributions of the excited states,
as determined from the observed shapes.  This point is discussed further
below.
The spectrograph was operated in dispersion-matched mode, in which
the intrinsic dispersion of the secondary beam (0.5$\%$) is compensated
by the last section of the spectrometer.  The targets of Be
and Au were 228 mg/cm$^2$ and 256 mg/cm$^2$ thick, respectively, chosen
to produce an energy straggling of about the same magnitude as the resolution
of the spectrograph. The overall resolution (including target thickness
effects) was measured to be 9 MeV/c full width at half maximum (FWHM)
in a direct-beam run with the target in place.  This run also gave the normalization for the beam flux, determined to be $5\times 10^3$ particles/s.

Time-of-flight information (over a distance of 70 m) combined with
the energy loss and total energy signal obtained with a segmented ion chamber
and a 5 cm thick plastic scintillator, respectively, were used
to identify and measure the yields of the fragments in the focal plane
of the spectrograph.  Two $x/y$ position-sensitive cathode-readout drift
chambers in the focal plane were used to determine the momentum and angle
information of the fragments.

\section{Experimental Results}

\subsection{Cross sections and spin assignments for $^{13}$B}

An energy spectrum of $\gamma$-rays in coincidence with
$^{13}$B fragments is shown in Fig. 1.  This spectrum has been
transformed into the projectile rest frame using the position of the
incident $\gamma$-ray measured in the NaI array, which allows for Doppler
correction on an event-by-event basis.  The solid curve is a fit to the
data obtained via Monte-Carlo simulation using the code GEANT \cite{gea94}.
The simulation took into account Doppler broadening, the distortion of the
shapes caused by the back transformation to the projectile rest frame, and
the
calculated $\gamma$-ray detection efficiencies.  Experimental efficiencies
measured with calibrated radioactive sources agreed with the simulation to
within 5$\%$.  The background, which presumably results from a combination 
of breakup reactions leaving the target in an excited state and secondary
interactions of neutrons with the detector and surrounding materials, 
was parameterized by a simple exponential dependence. Taking into account
the 3.48 MeV, 3.68 MeV and 4.13 MeV transitions in $^{13}$B,
the fit to the experimental spectrum is excellent (the $\chi^2$ per
degree of freedom N is 1.08).  The choice of these particular $\gamma$-ray
transitions is discussed below.  Eliminating the 4.13 MeV $\gamma$-ray from
the fit causes $\chi^2$/N to increase to 1.10
for 170 degrees of freedom, which
is marginally significant.  This indicates that all the important
transitions were accounted for and that there are no other significant
$\gamma$-rays between 1.5 and 3.5 MeV or above 4.2 MeV.

The level scheme of $^{13}$B \cite{ajz91} is presented in Fig. 2.
The ground state has J$^{\pi}={3 \over 2}^-$, resulting
from a [$1p_{3/2}$]$^{-1}$ proton configuration.  Below 4.5 MeV, three
negative parity and two positive parity excited states have been identified.
Although the energies and parities of these states are well
established from particle-transfer reaction studies, very little is 
known about their spins.  Based on the
$\gamma$-ray spectrum of Fig. 1, we are interested in the group of
states between 3.48 and 4.13 MeV.  

In the simplest model, the J$^{\pi}={2}^-$ ground state of $^{14}$B
results from the coupling of
a $2s1d$-shell neutron to the $^{13}$B ground state.  
Then, the
neutron removal from $^{14}$B could proceed by the removal of the 
$sd$-shell neutron to leave $^{13}$B in its ground state, by
removal of a neutron from the $1p_{1/2}$ orbital leaving $^{13}$B in
an excited state with J$^{\pi}=3/2^+,5/2^+$, or by
removal of a neutron from the $1p_{3/2}$ orbital leaving $^{13}$B in
a more highly excited state with 
J$^{\pi}=1/2^+,3/2^+,5/2^+$, or $7/2^+$.

This simple model is confirmed
by shell-model calculations in the $1p-2s1d$ model space with the
WBT and WBP residual interactions \cite{bro99}. Calculations were carried
out for the $^{14}$B ground state with a $(1p)^{-3} (sd)^1$
configuration (relative to a closed shell for for $^{16}$O), and for the
$^{13}$B spectrum with
$(1p)^{-3}(2s1d)^0$, $(1p)^{-4}(2s1d)^1$ and $(1p)^{-5}(2s1d)^2$
configurations.
(The total number of eigenstates are 5, 299 and 2973, respectively.)
The low-lying energy levels
obtained for these configurations, labeled by (0), (1) and (2),
respectively, are shown in Fig. 2. The spectroscopic factors
for the WBT interaction are given in Table 1 (the spectroscopic
factors obtained with WBP are similar to those of WBT). 
The higher levels, which
are reached by predominantly $1p_{3/2}$ removal, lie above the neutron
decay threshold of 4.9 MeV.
The
spectroscopic factors
leading to the (2) configurations are zero with our configuration
assumptions. 
All of the (0), (1)
and (2) configurations are allowed in the $^{11}$B(t,p)$^{13}$B
reaction which has been previously used to identify the $^{13}$B 
states \cite{ajz78}.
In principle, the (0) and (2) configurations can be mixed, which could
give some spectroscopic strength leading to negative parity states
around 4 MeV in excitation.
Our estimates within the 
mixed (0)+(2) model space indicate that 
the effect of this mixing leads to spectroscopic factors of
about 0.02 or less, and the effect of this mixing 
will be ignored here.

Partial cross sections obtained from the absolute $\gamma$-ray
intensities corrected by the computed efficiencies and the spectrometer
acceptance are presented in Table I.  The overall error in these cross
sections are of the order of 18$\%$, due mainly to systematic
errors in the fit to the experimental gamma-ray spectrum, and to the
uncertainties in the number of particles in the beam (12$\%$) and the
target thickness (5$\%$).

The calculated cross section for each final state of the
core fragment in a one-nucleon knockout reaction is given by:
\begin{equation}
\sigma(n) = \sum_j C^2S(j,n)\sigma_{sp}(j,B_n),
\end{equation}
where $C^2S(j,n)$ is the computed spectroscopic factor for the removed
nucleon with respect to a given core state and $\sigma_{sp}(j,B_n)$ is
the reaction cross section for the removal of a nucleon from a
single-particle state with total angular momentum $j$.
$B_n$ is the sum of the separation energy and excitation energy of the
state $n$. The single-particle cross sections were calculated by Tostevin
\cite{tos99} in a eikonal model assuming that there are two reaction
mechanisms involved in the knock-out reaction: (i) nucleon stripping in
which the halo nucleon interacts strongly with the target and leaves the
beam, and (ii) diffraction dissociation in which the nucleon moves
forward with essentially the beam velocity but away from the core.

The calculated cross sections for the states below 4.5 MeV are given in
Table I. The best agreement with experiment is achieved with the spin
assignments shown there. It should be noted that the evidence for the
existence of the 4.13 MeV transition in the experimental $\gamma$-ray
spectrum is marginal. Furthermore, the calculated strength for this
transition (Table I) does not take into account the configuration mixing
discussed above.  Overall, it appears that the combination of the
theoretical spectroscopic factors and computed single-particle cross
sections does a remarkable job of reproducing the experimental yields.

\subsection{Momentum distributions}

The momentum vector of the fragment at the target position after 
the reaction was reconstructed using the known magnetic fields 
and the positions and angles at the focal plane, and the ion-optics
code COSY \cite{ber93}. The momentum components transverse to the beam 
direction are sensitive to the reaction mechanism, while 
the longitudinal momentum component is relatively independent of the
details of the reaction mechanism as discussed by Orr, {\it et al.}
\cite{orr95}.  The longitudinal momentum distribution is obtained 
as the projection of the total momentum in the direction parallel to the beam.  
From the coincident $\gamma$-ray data, it is possible to 
generate the distribution for the case when
$^{13}$B is left in its ground state, by subtracting the contribution from the excited states 
from the singles spectrum after correcting for the 
$\gamma$-ray detection efficiencies.  The result of this subtraction process
is presented in Fig. 3(a) and corresponds to the removal of a single
neutron from the $sd$ shell in $^{14}$B due to a reaction with a $^9$Be target. 
Fig. 3(b) shows the longitudinal momentum distribution for the core-excited states, which is 
considerably broader.  

A more precise comparison can
be made through the use of the reaction theory.  Calculated longitudinal momentum distributions, 
also shown in Fig. 3, were determined within the framework of the eikonal model following the
procedure of Ref. \cite{han96}.  Since the dissociation products are formed at an impact
parameter  greater than $b_{min} = R_C + R_n$ (where $R_C$ and $R_n$ 
are the energy-dependent radii for the core and valence nucleon,
usually chosen to reproduce the measured interaction cross section) 
the black disc approximation can be used.  The cross section is 
then expressed in terms of the impact parameter as the 
one-dimensional Wigner transform of the wave function after the reaction, 
where the wave function profile is unity outside of a cutoff radius and zero inside.
The momentum distribution corresponding to the $^{13}$B core in its 
ground state, shown in Fig. 3(a),  agrees reasonably well with  the {\it l} = 0 
curve of the eikonal model calculation, which represents the removal of an $s$-wave valence 
neutron from the $sd$ shell in $^{14}$B. Similar 
results have been obtained for $^{11}$Be \cite{aum99} and $^{15}$C \cite{mad00}.
The width obtained for the distribution is $55~\pm~2$  MeV/c FWHM, after correcting for 
the 9 MeV/c experimental momentum resolution.  This width results from a Lorentzian fit to the
experimental data and is in excellent agreement with the value of 56 MeV/c FWHM displayed by
the theoretical  {\it l} = 0 curve in the lab. system.  The disagreement in the tails of the
distribution suggests that there may be a contribution from the {\it l} = 2 distribution and
possibly other effects due to the acceptance of the spectrometer, approximations
in the reaction model used to analyze the data, or unidentified contributions
to the line shape \cite{aum99}.

Spectroscopic factors of 0.306 and 0.662, respectively, 
are obtained from the shell model calculation \cite{bro99} for the removal 
of a valence neutron from the $1d_{5/2}$ and $2s_{1/2}$
states in $^{14}$B, which indicates that a non-negligible {\it l} = 2
admixture is expected.  Individual cross sections for both 
configurations have been derived by fitting the experimental distribution with  a linear  
combination of {\it l} = 0 and {\it l} = 2 shapes.  The best fit, shown in Fig. 4, implies
that the removal of a valence  neutron from the $2s_{1/2}$ ($1d_{5/2}$) orbital in the $^{14}$B
ground state accounts for 89$\pm 3\%$ (11$\pm 3\%$) of the total
yield.  The agreement with cross sections derived from the eikonal 
model and the theoretical spectroscopic factors is very good (Table I),
as is the quality of the fit to the experimental data shown in Fig. 4.  

The core-excited longitudinal momentum distribution, shown in Fig. 3(b),
was obtained from the $\gamma$-ray-coincident $^{13}$B yield after subtracting
a background distribution derived by gating on the exponential part of
the $\gamma$-ray spectrum at high energy and normalizing to the
extrapolated background under the peaks.  The background computed in this
way constituted 26$\%$ of the coincident data in the region of the
three identified $\gamma$-ray transitions.  The 
core-excited distribution has a shape that agrees well with that 
expected for {\it l } = 1, confirming the expectation of the removal of a $1p_{1/2}$
neutron.   The experimental width, obtained from a 
Lorentzian fit to the central part of the distribution, 
is $135~\pm~15$ MeV/c FWHM. The calculated width in the 
lab. frame of the theoretical {\it l} = 1 curve is 144 MeV/c.  The
distributions computed for $p$ and $d$ wave breakup are broader than 
the actual momentum acceptance of the spectrometer.  A (small) correction 
to the integrated yield due to this limited acceptance  
has been applied to the experimental cross sections given in
Table I. 

\subsection{Coulomb breakup}

The longitudinal momentum distribution obtained in the
one-neutron breakup reaction of $^{14}$B on a $^{197}$Au target is
shown in Fig. 5.   The $\gamma$-ray coincident yield has been subtracted
using the same procedure as described in the previous section.  In this case, the $\gamma$-ray spectrum was essentially equal to the exponential background distribution, and no core-excited transitions were observed.
The width obtained from a gaussian fit (dashed curve in Fig. 5) to the experimental longitudinal momentum distribution is $59\pm3$ MeV/c 
(after correcting for the experimental resolution). This width is larger 
than the value of $48\pm3$ MeV/c measured by Bazin, {\it et al.} \cite{baz98} for Coulomb breakup on a tantalum target at 86 MeV/A.  However, it agrees  
within the errors with that extracted from the $^{13}$B 
ground-state longitudinal momentum distribution for the Be target in
the present work.  The integrated cross section derived for Coulomb breakup of $^{14}$B on $^{197}$Au (leaving the $^{13}$B core fragment in the ground-state) is $638\pm45$ mb, much larger than the yield quoted in Ref.\cite{baz98}.
It was suggested there that the smaller cross section reflected a
quenching of the soft dipole strength that implied ``normal" nuclear
structure for $^{14}$B.  However, a possible experimental problem
due to the restricted angular acceptance of the A1200 spectrometer
was acknowledged.  The present observations of a broader momentum
distribution coupled with a larger cross section suggest that this
was indeed the case.

The Coulomb breakup cross section was calculated using a Yukawa potential
with finite-size corrections \cite{anne94}.  The result, 543 mb,
is less than our measurement.  However, the yield
calculated in Ref. \cite{baz98} using a Woods-Saxon potential was 50$\%$
larger than that from the Yukawa form.  Our data suggest that the
actual situation is intermediate between these two approximations.  The
Coulomb contribution to our Be-target data was also computed and found
to be negligibly small (2.3 mb).  The solid curve in Fig. 5 shows the
longitudinal momentum distribution for the Au target predicted in the model of Ref. \cite{anne94},
normalized to the experimental data.  The width of this Lorentzian
function compares well with experiment, but the agreement in the 
tails of the distribution is unsatisfactory.

\subsection{Neutron halo in $^{14}$B}

Zhongzhou, {\it et al.} \cite{zho97} have predicted an inversion of the $1d_{5/2}$ and $2s_{1/2}$ 
orbitals for $^{14}$B within the framework
of a non-linear relativistic mean-field calculation.  This is in
agreement with the shell-model calculation cited above and with
our observation of a dominant {\it l} = 0 component
in the ground state of this nucleus.  Their calculation shows
that, under these conditions, $^{14}$B is a neutron halo nucleus.
On the other hand, measurements of the interaction cross sections at
relativistic energies for particle-stable B isotopes ranging from mass 
8-15, by Tanihata, {\it et al.} \cite{tan88}, suggested that their effective RMS
radii are practically constant, which does not support the halo hypothesis.  Measurements 
at intermediate energy,
analyzed by Liatard, {\it et al.} \cite{lia90}, gave larger RMS radii in
all cases than those of Ref. \cite{tan88}, together with a significant
mass dependence and an anomalously large radius for $^{14}$B.  The
radius difference is not by any means as large as in the case of
$^{11}$Li, for example, but this is not unexpected given the fact that
the valence neutron is more tightly bound in $^{14}$B.  Of course, the RMS radius extracted 
from an interaction cross section is not the only signature for halo structure \cite{orr97,ris97,gui00},
and the narrow longitudinal momentum distribution and large cross section for dissociation of the ground state of $^{14}$B observed
in the present experiment strongly argue for halo structure in this system.

\section{Summary}

In this experiment, we have measured the longitudinal momentum of the 
$^{13}$B core fragment in one-neutron knockout from 
$^{14}$B, on both $^9$Be and $^{197}$Au targets.  The contribution from 
core excitation was isolated by measuring the excited $^{13}$B fragments 
in coincidence with $\gamma$-rays 
from their decay.  Comparison of the observed intensities
of core-excited $\gamma$-ray transitions with a shell-model 
calculation, using an eikonal reaction theory, allowed us to
make spin assignments for two positive-parity excited states in
$^{13}$B.  The longitudinal momentum distributions in coincidence with the
positive-parity core-excited transitions are consistent with knockout from a
$1p_{3/2}$ state, as expected.  An excellent fit to the distribution obtained when
the $^{13}$B core remains in its ground state is also provided by 
the shell-model calculation, which predicts an admixture
of $2s_{1/2}$ and $1d_{5/2}$ neutron knockout.  The experimental
data imply an 89$\pm 3\%$ {\it l} = 0 component, in agreement 
with the shell-model result.  The Coulomb breakup cross section,
measured with the $^{197}$Au target, is consistent with expectations
for a weakly-bound $2s_{1/2}$ neutron.  The width of the longitudinal
momentum distribution obtained with the $^{197}$Au target ($59\pm3$ MeV/c)
agrees with that obtained using the $^{9}$Be target after the
core-excited component is subtracted.
 
The results of the present experiment lend very strong support to
the idea that $^{14}$B is a ``neutron-halo" system, the first
odd-odd nucleus to display this structure.  It also appears that,
with the sole exception \cite{baz98} of $^{17}$C at N=11, all of the 
lowest-mass, particle-stable isotones from N=7-13 are halo nuclei.
It therefore seems that re-investigation of the structure of
$^{17}$C is in order.

\section{ acknowledgments }

 The first author (V.G.)  was financially 
supported by FAPESP (Funda\c c\~ao de 
Amparo a Pesquisa do Estado de S\~ao Paulo - Brazil) while on leave 
from the UNIP (Universidade Paulista). 
This work was supported by the National Science Foundation under 
Grants No. PHY94-02761, PHY99-01133, PHY95-14157, and PHY96-05207.

\end{multicols}

\vspace{1cm}

\begin{table}
\caption{Partial cross sections in mb for each of the final states populated
in $^{13}$B.  The spin assignments for each level (third
column) are discussed in the text. The fourth and fifth columns are the 
orbital angular momentum of the valence neutron
and the corresponding spectroscopic factor,
respectively.  The next three columns
give the partial cross sections for stripping and diffractive
breakup in the eikonal model of Ref. \protect\cite{tos99}, and
the theoretical value for the cross section which is the sum of the
two components multiplied by the corresponding spectroscopic factor.
The experimental value is given in the last column. }
\begin{center}
\begin{tabular}{ccccccccc}
E$^{exp}$(MeV) & E$^{theo}(MeV) $ & J$^{\pi}$ & {\it l} &  $S$  & $\sigma^{strip}_{SP}$ &
$\sigma^{diff}_{SP}$ &  $\sigma^{theo}$ & $\sigma^{exp}$\\
\hline
0.00 & 0.00 & ${3\over 2}^-$ & 0 & 0.662 & 82.4 & 53.9 & 90.2 &  113 (15)~${^a}$ \\
     &   &                           & 2 & 0.306 & 35.0 & 16.5 & 15.8 &
~14 (~3)~${^a}$\\
     &  3.43 & ${1\over 2}^+$ & 1 & 0.092 & 24.1 &  9.7  & 3.1 &  \\
3.48 &  4.00 & ${3\over 2}^+$ & 1 & 0.407 & 24.1 &  9.7  & 14.0 & 18 (3) \\
3.68 & 4.06 & ${5\over 2}^+$ & 1 & 0.886  & 23.5 &  9.4  & 29.2 & 30 (5)  \\
     &   &                           &    &            &         &         &
&       \\
4.13 & &  &  & &  &   &  & 1.2 (1.2)  \\
     & 4.80 & ${1\over 2}^-$ & 2 &0.0005 & 22.3 &  8.0  & 0.014 &   \\
\end{tabular}
\end{center}
{\footnotesize $^{(a)}$ Cross section obtained by deconvolution of the ${\it l}$=0 and 
${\it l}$=2
contributions to the longitudinal momentum distribution.}
\end{table}

\begin{figure}
\caption{Doppler corrected energy spectrum of the $\gamma$-rays measured
in coincidence with a $^{13}$B fragment in the NaI array. The solid line is
 the fit to the experimental data. The fit corresponds to an independent
normalization  of
the simulated response functions for each individual $\gamma$-ray,
 indicated in the figure by the gray curves.
The background is parametrized by an exponential dependence.}
\end{figure}

\begin{figure}
\caption{The level scheme of $^{13}$B. The experimental data are from
Ref. \protect\cite{ajz91} (and references therein), and the theoretical
predictions are from a shell model calculation. The values given in parentheses are
the numbers of neutrons excited to the $sd$ shell.
The dashed lines correspond to the assignments discussed in the text.}
\end{figure}

\begin{figure}
\caption{The longitudinal momentum distribution of the
 $^{13}$B core fragment, in the laboratory frame,  from one-neutron
knockout  reaction from $^{14}$B incident on a $^9$Be target. {\bf (a)} Longitudinal
momentum distribution corresponding to the $^{13}$B core in its
ground state.  {\bf (b)} The same, but for $^{13}$B core-excited
states. The error bars indicate only the statistical uncertainties.
 The thick solid, dashed, and thin solid curves correspond to
the momentum distributions calculated in the eikonal model for the removal
of a neutron in an $s$, $p$, or $d$ orbital, respectively. }
\end{figure}

\begin{figure}
\caption{The longitudinal momentum distribution with the $^{13}$B core fragment
in its ground state. The thick solid line is the result of a fit with a
linear combination
of the {\it l}~=~0 and {\it l}~=~2 shapes from the
eikonal model. The dashed curve and thin solid
curve are the individual contributions for {\it l}~=~0 and {\it l}~=~2,
respectively.}
\end{figure}

\begin{figure}
\caption{The longitudinal momentum distribution with the
 $^{13}$B core fragment in its ground state for the Coulomb breakup
of $^{14}$B on a $^{197}$Au target. The dashed curve is a gaussian 
fit having a width of $60\pm3$ MeV/c FWHM.  The solid curve is
the result of a calculation using a Yukawa potential with
finite-size corrections \protect\cite{anne94}.}
\end{figure}

\end{document}